\documentclass[aps,superscriptaddress,twocolumn,amsmath,amssymb,preprintnumbers,showpacs]{revtex4}
\usepackage{amssymb}
\usepackage{dcolumn}
\usepackage{bm}
\usepackage{graphicx}
\usepackage{booktabs}
\usepackage{multirow}
\usepackage{subfig}

\begin{document}

\newcommand*{\PKU}{School of Physics and State Key Laboratory of Nuclear Physics and
Technology, Peking University, Beijing 100871,
China}\affiliation{\PKU}
\newcommand*{\CHEP}{Center for High Energy Physics, Peking University, Beijing 100871, China}\affiliation{\CHEP}

\title{On the CP-violating phase $\delta_{\rm CP}$ in fermion mixing matrices}

\author{Xinyi Zhang}\affiliation{\PKU}
\author{Bo-Qiang Ma}\email{mabq@pku.edu.cn}\affiliation{\PKU}\affiliation{\CHEP}

\begin{abstract}
The recent established large $\theta_{13}$ in neutrino mixing
provides an optimistic possibility for the investigation of the CP
violation, therefore it is necessary to study the CP-violating phase
$\delta_{\rm CP}$ in detail. Based on the maximal CP violation
hypothesis in the original Kobayashi-Maskawa (KM) scheme of neutrino
mixing matrix, i.e., $\delta_{\rm KM}=90^\circ$, we calculate
$\delta_{\rm CK}$ for both quarks and leptons in the Chau-Keung (CK)
scheme of the standard parametrization and find that
$\delta^{\mathrm{quark}}_{\mathrm{CK}}=(68.62^{+0.89}_{-0.81})^\circ$
and
$\delta^{\mathrm{lepton}}_{\mathrm{CK}}=(85.39^{+4.76}_{-1.82})^\circ$,
provided with three mixing angles to be given. We also examine the
sensitivity of $|V_{ij}|$ and $|U_{ij}|$ to $\delta_{\rm CK}$ and
$\delta_{\rm KM}$. As a convention-independent investigation, we
discuss the $\Phi$ matrix, which has elements correspond to angles
of the unitarity triangles. We demonstrate the $\Phi$ matrices for
both quark and lepton sectors and discuss the implications as well
as the variations of the $\Phi$ matrix elements with $\delta_{\rm
CP}$.

\end{abstract}


\maketitle

\section{INTRODUCTION}
The oscillation of neutrinos have been verified for more than a
decade, and such a phenomenon can be described by the standard model
of particle physics (SM) as the misalignment of the flavor
eigenstates with the mass eigenstates, as shown explicitly in the
lagrangian of the charged current (CC) interaction
\begin{eqnarray}
L&=&-\frac{g}{\sqrt{2}}U^\dagger_L \gamma^\mu V_{\rm CKM} D_L
W^+_\mu
\nonumber \\
&&-\frac{g}{\sqrt{2}}E^\dagger_L \gamma^\mu U_{\rm PMNS}N_LW^-_\mu +
h.c.,\label{lagrangian}
\end{eqnarray}
where
\begin{eqnarray}
&U_L=(u_L,c_L,t_L)^T; \ \
&D_L=(d_L,s_L,b_L)^T;\nonumber\\
&E_L=(e_L,\mu_L,\tau_L)^T; \ \ &N_L=(\nu_1,\nu_2,\nu_3)^T.
\end{eqnarray}
We discuss the situation with only three generations, while
additional generation can also be collaborated in the lagrangian. In
Eq.(\ref{lagrangian}), $V_{\rm CKM}$, namely the
Cabibbo-Kobayashi-Maskawa~(CKM) matrix~\cite{CKM1,CKM2}, is the
mixing matrix describing the mixing between different generations of
quarks, and correspondingly $U_{\rm PMNS}$, the
Pontecorvo-Maki-Nakagawa-Sakata (PMNS) matrix~\cite{PMNS}, describes
the misalignment of the flavor eigenstates with the mass eigenstates
of leptons. By choosing the mass matrix of charged leptons to be
diagonal, the PMNS matrix represents the neutrino mixing, therefore
we can also call it the neutrino mixing matrix.

Both the CKM matrix and the PMNS matrix are unitary matrices that
can be parameterized by three Euler angles representing three
rotations in certain planes and one Dirac phase angle representing
the CP violation. If the neutrinos are of Majorana type, two
additional Majorana phases are needed to fully determine the mixing
matrix. While the Majoranna phases do not manifest themselves in the
oscillation, we discuss the Dirac phase only. This kind of
parametrization, which can be referred to as the angle-phase
parametrization, has the freedom of arranging the orders of these
three rotations. Of the twelve ways to do the product, only nine are
independent and the standard parametrization, i.e., the Chau-Keung
(CK) scheme~\cite{CK} adopted by Particle Data
Group~\cite{PDG1996,PDG2008,pdg2010}, is one of the
nine~\cite{9/12,Zheng10,Zhang:2012xu}.

Recent progress on the measurement of the smallest neutrino mixing
angle $\theta_{13}$ by the Daya-Bay~\cite{An:2012eh} and RENO
collaborations~\cite{Ahn:2012nd} has established a non-zero and
relatively large value. This progress might be considered as a
signal of the era of precise measurement of neutrino oscillation as
well as an optimistic possibility for future measurements of the CP
violating phase $\delta_{\rm CP}$ in the neutrino mixing. In fact, a
non-zero $\theta_{13}$ was indicated by various experiments, i.e.,
the T2K, MINOS and Double-Chooz collaborations since last
year~\cite{Abe:2011sj,Adamson:2011qu,Abe:2011fz}.

To accommodate the experimental data of neutrino mixing, a certain
parametrization should be adopted. As mentioned before, the
Chau-Keung (CK) scheme is adopted as the standard one, and the
mixing angles in this scheme are directly related to the observed
oscillation probabilities. Since all the parameterizations are
equivalent to each other mathematically and other schemes may still
have some advantages in phenomenological analysis as well as model
building, it is meaningful to explore schemes other than the
standard one for the possibility to find some clues towards a better
understanding of fermion properties. For example, the original
Kobayashi-Maskawa (KM) scheme~\cite{CKM2} allows for almost a
perfect maximal CP violation of the quark mixing, i.e., the CP
violating phase
$\delta^{\mathrm{quark}}_{\mathrm{KM}}=90^\circ$~\cite{koide,Koide:2008yu,boomerang,Li:2010ae,qinnan,Ahn:2011it},
whereas in the standard parametrziation
$\delta^{\mathrm{quark}}_{\mathrm{CK}}=68.8^\circ$\cite{pdg2010},
which deviates from the maximal CP violation. This arises naturally
the following questions: Is there a maximal CP violation in the
lepton sector? What happens if
$\delta^{\mathrm{lepton}}_{\mathrm{KM}}=90^\circ$? Is there any
reason behind the hypothesis? As there is no experimental
information on $\delta^{\mathrm{lepton}}_{\mathrm{CP}}$ yet, we take
$\delta^{\mathrm{lepton}}_{\mathrm{KM}}=90^\circ$ as an Ansatz, to
see what can we get for neutrino mixing.

Cautions should be taken when talking about the maximal CP
violation, because historically, the notion refers to a maximized
$J$~\cite{Dunietz:1985uy}, i.e., the Jarlskog
invariant~\cite{jarlskog}. The conception we adopt here refers to
the case that the CP-violating phase equals to $90^\circ$, which
guaranties the term $\sin\delta$ in its maximal value as such term
always shows up in the Jarlskog invariant~\cite{Koide:2008yu}.

We make our Ansatz of a maximal CP violation
$\delta_{\mathrm{KM}}^{\mathrm{lepton}}=90^\circ$ in the lepton
sector based on all the similarities shared by quarks and leptons
and the fact that there is no information on the CP-violating phase
experimentally now. Based on the Ansatz, we can work out all the
elements of the neutrino mixing matrix together with a prediction of
the CP-violating phase $\delta_{\mathrm{CK}}^{\mathrm{lepton}}$, as
has been shown in Ref.~\cite{Zhang:2012ys} briefly. In
Section~\ref{sec:predict}, we firstly perform a replaying procedure
in the quark sector as a test for
$\delta^{\mathrm{quark}}_{\mathrm{KM}}=90^\circ$ as well as an
exercise for the method. Then we take
$\delta^{\mathrm{lepton}}_{\mathrm{KM}}=90^\circ$ as an Ansatz to
provide a prediction of $\delta^{\mathrm{lepton}}_{\mathrm{CK}}$. In
Section~\ref{sec:moduli}, we examine the dependence of the mixing
matrix elements on the CP-violating phase. As there is a rephasing
freedom in the mixing matrix, we also discuss the
convention-independent $\Phi$ matrix in Section~\ref{sec:phi}, where
we provide predictions for all of the unitarity triangles of
neutrinos. Section~\ref{sec:sum} serves for some discussions and
conclusions.

\section{\label{sec:predict} The CP-VIOLATING PHASES $\delta_{\mathrm{CK}}$ IN QUARK AND LEPTON SECTORS }

\subsection{Reproduction of $\delta_{\mathrm{CK}}$ in the quark sector}

The four parameters needed to determine the CKM matrix have been
measured to high precision. Using the global fit result of the four
Wolfenstein parameters~\cite{pdg2010}, we can get three mixing
angles together with the CP-violating phase in any angle-phase
parametrization. In the following we perform a replay to obtain the
CP-violating phase $\delta^{\mathrm{quark}}_{\mathrm{CK}}$ in the
standard parametrization (i.e., the CK scheme) provided with three
mixing angles to be given together with a maximal CP violation
hypothesis in the KM scheme. Our purpose is to check the procedure
and then to make prediction of the CP-violating phase
$\delta^{\mathrm{lepton}}_{\mathrm{CK}}$ for the lepton sector. Our
calculation is to obtain $\delta^{\mathrm{quark}}_{\rm CK}$ from
$\delta^{\mathrm{quark}}_{\rm KM}=90^\circ$, i.e.,  we assume that
$\delta^{\mathrm{quark}}_{\rm CK}$ is unknown. The mixing angles in
the standard parametrization are deduced from the Wolfenstein
parameters,
\begin{eqnarray}
s_{12}=\lambda=0.2253\pm0.0007;\nonumber\\
s_{23}=A \lambda^2=0.0410^{+0.0011}_{-0.0008};\nonumber\\
s_{13}=|A \lambda^3(\rho+i \eta)|=0.0035^{+0.0002}_{-0.0001},
\end{eqnarray}
in which $s_{ij}=\sin\theta_{ij}$ and $c_{ij}=\cos\theta_{ij}$
($i,j=1,2,3$). From the expression of the Chau-Keung (CK) scheme
\begin{widetext}
\begin{eqnarray}
V_{\rm CK}&=&\left(
  \begin{array}{ccc}
    1  & 0     & 0         \\
    0  & c_{23}  & s_{23} \\
    0  & -s_{23} & c_{23} \\
  \end{array}
\right)\left(
  \begin{array}{ccc}
    c_{13}                & 0 & s_{13}e^{-i\delta_{\rm CK}} \\
    0                     & 1 & 0          \\
    -s_{13}e^{i\delta_{\rm CK}} & 0 & c_{13} \\
  \end{array}
\right)\left(
  \begin{array}{ccc}
    c_{12}  & s_{12} & 0 \\
    -s_{12} & c_{12} & 0 \\
    0       & 0      & 1 \\
  \end{array}
\right)\nonumber\\
&=&\left(
\begin{array}{ccc}
c_{12}c_{13} & s_{12}c_{13} & s_{13}e^{-i\delta_{\rm CK}}         \\
-s_{12}c_{23}-c_{12}s_{23}s_{13}e^{i\delta_{\rm CK}} & c_{12}c_{23}-s_{12}s_{23}s_{13}e^{i\delta_{\rm CK}} & s_{23}c_{13} \\
s_{12}s_{23}-c_{12}c_{23}s_{13}e^{i\delta_{\rm CK}} & -c_{12}s_{23}-s_{12}c_{23}s_{13}e^{i\delta_{\rm CK}} & c_{23}c_{13}\\
\end{array}\label{CK}
\right),
\end{eqnarray}
\end{widetext}
we can get the moduli of five CKM matrix elements
\begin{eqnarray}
|V_{ud}|&=&c_{12}c_{13}=0.9743\pm0.0002;\nonumber\\
|V_{us}|&=&s_{12}c_{13}=0.2253\pm0.0007;\nonumber\\
|V_{ub}|&=&s_{13}=0.0034^{+0.0002}_{-0.0001};\nonumber\\
|V_{cb}|&=&s_{23}c_{13}=0.0410^{+0.0011}_{-0.0008};\nonumber\\
|V_{tb}|&=&c_{23}c_{13}=0.9992^{+0.00005}_{-0.00003}.
\end{eqnarray}
Substituting the five elements values into the expression of the
Kobayashi-Maskawa (KM) scheme,
\begin{widetext}
\begin{eqnarray}
V_{\rm KM}&=\left(
  \begin{array}{ccc}
    1 & 0   & 0    \\
    0 & c_2 & -s_2 \\
    0 & s_2 & c_2  \\
  \end{array}
\right)\left(
  \begin{array}{ccc}
    c_1 & -s_1 & 0 \\
    s_1 & c_1  & 0 \\
    0   & 0    & e^{i\delta_{\rm KM}} \\
  \end{array}
\right)\left(
  \begin{array}{ccc}
    1 & 0   & 0    \\
    0 & c_3 & s_3  \\
    0 & s_3 & -c_3 \\
  \end{array}
\right)\nonumber\\
&=\left(
\begin{array}{ccc}
c_1     & -s_1c_3                     & -s_1s_3           \\
s_1c_2  & c_1c_2c_3-s_2s_3e^{i\delta_{\rm KM}} & c_1c_2s_3+s_2c_3e^{i\delta_{\rm KM}} \\
s_1s_2  & c_1s_2c_3+c_2s_3e^{i\delta_{\rm KM}} &
c_1s_2s_3-c_2c_3e^{i\delta_{\rm KM}}
\end{array}\label{KM}
\right),
\end{eqnarray}
\end{widetext}
together with the input $\delta^{\mathrm{quark}}_{\rm KM}=90^\circ$,
we can get the mixing angles of this scheme,
\begin{eqnarray}
\theta_1&=&(13.02\pm0.04)^\circ;\nonumber\\
\theta_2&=&(2.19^{+0.06}_{-0.04})^\circ;\nonumber\\
\theta_3&=&(0.88\pm0.04)^\circ.
\end{eqnarray}
The corresponding trigonometric functions are
\begin{eqnarray}
\sin\theta_1&=&0.2253\pm0.0007;\nonumber\\
\cos\theta_1&=&0.9743\pm0.0002;\nonumber\\
\sin\theta_2&=&0.0382^{+0.0011}_{-0.0008};\nonumber\\
\cos\theta_2&=&0.9993^{+0.00004}_{-0.00003};\nonumber\\
\sin\theta_3&=&0.0150^{+0.0007}_{-0.0006};\nonumber\\
\cos\theta_3&=&0.9999\pm0.00001.
\end{eqnarray}
Now we have values for all the four parameters in
Eq.(\ref{KM}), so we can get all the moduli of the CKM matrix,
\begin{widetext}
\begin{align}
|V_{\rm CKM}|= \left(
  \begin{array}{ccc}
   0.9743\pm0.0002           & 0.2253\pm0.0007            & 0.0035^{+0.0002}_{-0.0001}\\
   0.2252\pm0.0007           & 0.9735\pm0.0002            & 0.0410^{+0.0011}_{-0.0008} \\
   0.0086^{+0.0003}_{-0.0002}& 0.0403^{+0.0011}_{-0.0008} & 0.9992^{+0.00005}_{-0.00003}
  \end{array} \right),\label{CKMprediction}
\end{align}
\end{widetext}
which is consistent with the global fit result in Ref.~\cite{pdg2010} as expected.

As a useful quantity describing the magnitude of CP violation, the Jarlskog invariant~\cite{jarlskog} can be calculated in the Kobayashi-Maskawa (KM) scheme,
\begin{eqnarray}
J^{\mathrm{quark}}_{\rm
KM}&=&\frac{1}{8}\sin\theta_1\sin2\theta_1\sin2\theta_2\sin2\theta_3\sin\delta^{\mathrm{quark}}_{\rm
KM}\nonumber\\&=&(2.90^{+0.19}_{-0.15})\times10^{-5}.
\end{eqnarray}
Thus we can make a reproduction of the CP phase in the Chau-Keung
(CK) scheme by solving the equation
$J^{\mathrm{quark}}_{\rm
KM}=J^{\mathrm{quark}}_{\rm CK}$, where
\begin{eqnarray}
J^{\mathrm{quark}}_{\rm
CK}=\frac{1}{8}\cos\theta_{13}\sin2\theta_{12}\sin2\theta_{23}\sin2\theta_{13}\sin\delta^{\mathrm{quark}}_{\rm
CK}.~~
\end{eqnarray}
We get
\begin{eqnarray}
\delta^{\mathrm{quark}}_{\mathrm{CK}}=(68.62^{+0.89}_{-0.81})^\circ,
\end{eqnarray}
which is in accordance with the result extracted from the
Wolfenstein parameters~\cite{pdg2010}, which indicate
\begin{eqnarray}
\delta^{\mathrm{quark}}_{\mathrm{CK}}=(68.81^{+3.30}_{-2.17})^\circ.
\end{eqnarray}

\subsection{Prediction of $\delta_{\mathrm{CK}}$ in the lepton sector}
We have seen that the hypothesis of the maximal CP violation can
reproduce a reasonable $\delta^{\mathrm{quark}}_{\mathrm{CK}}$.
Under the same procedure, together with the Ansatz of a maximal CP
violation in the lepton sector, i.e.,
$\delta^{\mathrm{lepton}}_{\mathrm{KM}}=90^\circ$ for the KM-scheme
of mixing matrix, we can make a prediction of
$\delta^{\mathrm{lepton}}_{\mathrm{CK}}$~\cite{Zhang:2012ys}.

We adopt the global fit of neutrino mixing angles based on previous
experimental data including T2K and MINOS
experiments~($1\sigma~(3\sigma)$)~\cite{global}
\begin{eqnarray}
\sin^2\theta_{12}&=&0.312^{+0.017}_{-0.016}(^{+0.052}_{-0.047});\nonumber\\
\sin^2\theta_{23}&=&0.42^{+0.08}_{-0.03}(^{+0.22}_{-0.08}),
\end{eqnarray}
for our input of $\theta_{12}$ and $\theta_{23}$.
While for $\theta_{13}$, it is reasonable to make use of all the recent data showed in Table~\ref{tab:theta13}.
\begin{table*}
\caption{\label{tab:theta13} The recent experimental data on $\theta_{13}$}
\begin{ruledtabular}
 \begin{tabular}{ccc}
   \toprule
 Experimental collaboration & Data on $\theta_{13}$  &  \\
 \hline
 T2K
 &$0.03(0.04)<\sin^22\theta_{13}<0.28(0.34)$,~~NH(IH)
 &$\sin^22\theta_{13}=0.11\pm0.17$\footnote{Taking the latest results into account, we adopt the NH case with a symmetrized error range $\pm0.17$.}\\
 MINOS
 &$2\sin^2\theta_{23}\sin^22\theta_{13}=0.041^{+0.047}_{-0.031}(0.079^{+0.071}_{-0.053})$,~~NH(IH)
 &$\sin^22\theta_{13}=0.049\pm0.075$\footnote{We take the NH case in accordance with the former one and extract $\theta_{13}$ with $\theta_{23}$ taking its global fit value.} \\
 Double Chooz&$\sin^22\theta_{13}=0.086\pm0.041~(\mathrm{stat})\pm0.030~(\mathrm{syst})$
 &$\sin^22\theta_{13}=0.086\pm0.051$\footnote{The $1\sigma$ deviation is
 estimated by $\sigma^2=\sigma_{\mathrm{stat}}^2 +
 \sigma_{\mathrm{syst}}^2$.} \\
 Daya Bay
 &$\sin^22\theta_{13}=0.092\pm0.016~(\mathrm{stat})\pm0.005~(\mathrm{syst})$ &$\sin^22\theta_{13}=0.092\pm0.017$\footnotemark[3] \\
 RENO
 &$\sin^22\theta_{13}=0.113\pm0.013~(\mathrm{stat})\pm0.019~(\mathrm{syst})$ &$\sin^22\theta_{13}=0.113\pm0.023$\footnotemark[3] \\
 \hline
 The weighted average of $\theta_{13}$ & &$\sin^22\theta_{13}=0.097\pm0.013$\\
    \bottomrule
 \end{tabular}
\end{ruledtabular}
\end{table*}
The resulting $\theta_{13}$ is
\begin{eqnarray}
\sin^22\theta_{13}=0.097\pm0.013.
\end{eqnarray}
Explicitly, the three mixing angles in our input are
\begin{eqnarray}
\theta_{12}&=&(33.96^{+1.03}_{-0.99}(^{+3.22}_{-2.91}))^\circ;\nonumber\\
\theta_{23}&=&(40.40^{+4.64}_{-1.74}(^{+12.77}_{-4.64}))^\circ;\nonumber\\
\theta_{13}&=&(9.07\pm0.63(\pm1.89))^\circ.
\end{eqnarray}
The moduli of five matrix elements are
\begin{eqnarray}
|U_{e1}|&=&c_{12}c_{13}=0.819\pm0.010;\\
|U_{e2}|&=&s_{12}c_{13}=0.552^{+0.015}_{-0.014};\\
|U_{e3}|&=&|s_{13}|=0.158\pm0.011;\\
|U_{\mu3}|&=&s_{23}c_{13}=0.640^{+0.061}_{-0.023};\\
|U_{\tau3}|&=&c_{23}c_{13}=0.752^{+0.052}_{-0.019}.
\end{eqnarray}
With these five moduli, together with an Ansatz of maximal CP
violation $\delta_{\mathrm{KM}}^{\mathrm{lepton}}=90^\circ$, we can
get the mixing angles in the KM parametrization
\begin{eqnarray}
\theta_1&=&(35.01^{+1.02}_{-0.96})^\circ;\nonumber\\
\theta_2&=&(39.85^{+5.20}_{-1.95})^\circ;\nonumber\\
\theta_3&=&(15.96\pm1.14)^\circ.
\end{eqnarray}
The corresponding trigonometric functions are
\begin{eqnarray}
\sin\theta_1&=&0.574^{+0.015}_{-0.014},\quad\cos\theta_1=0.819\pm0.010;\\
\sin\theta_2&=&0.641^{+0.070}_{-0.026},\quad\cos\theta_2=0.768^{+0.058}_{-0.022};\\
\sin\theta_3&=&0.275\pm0.019,\quad\cos\theta_3=0.961\pm0.005.\label{KMpara}
\end{eqnarray}
Thus we have all the parameters in Eq.(\ref{KM}). Then we can get all the moduli of the PMNS matrix, which
is
\begin{align}
|U_{\rm PMNS}|= \left(
  \begin{array}{ccc}
   0.819\pm0.010            & 0.552^{+0.015}_{-0.014}  &  0.158\pm0.011   \\
   0.440^{+0.035}_{-0.016}  & 0.630^{+0.039}_{-0.016}  &  0.640^{+0.061}_{-0.023} \\
   0.368^{+0.041}_{-0.018}  & 0.547^{+0.045}_{-0.018}  &  0.752^{+0.052}_{-0.019}
  \end{array} \right).\label{PMNSprediction}
\end{align}
The corresponding Jarlskog invariant is
\begin{eqnarray}
J^{\mathrm{lepton}}_{\rm
KM}&=&\frac{1}{8}\sin\theta_1\sin2\theta_1\sin2\theta_2\sin2\theta_3\sin\delta^{\mathrm{lepton}}_{\rm
KM}\nonumber\\&=&0.035^{+0.003}_{-0.002}.
\end{eqnarray}
As the same procedure used in the quark sector, we can give our prediction of the CP-violating phase in the Chau-Keung (CK) scheme by solving the equation $J^{\mathrm{lepton}}_{\rm CK}=J^{\mathrm{lepton}}_{\rm
KM}$, where
\begin{eqnarray}
J^{\mathrm{lepton}}_{\mathrm{CK}}=\frac{1}{8}\cos\theta_{13}\sin2\theta_{12}\sin2\theta_{23}\sin2\theta_{13}\sin\delta^{\mathrm{lepton}}_{\mathrm{CK}}.~~
\end{eqnarray}
The resulting $\delta^{\mathrm{lepton}}_{\mathrm{CK}}$ is
\begin{eqnarray}
\delta^{\mathrm{lepton}}_{\mathrm{CK}}=(85.39^{+4.76}_{-1.82})^\circ.
\end{eqnarray}
Note that corresponding to our input of the mixing angles, $J^{\mathrm{lepton}}$ is in the range
\begin{eqnarray}
0\leq J^{\mathrm{lepton}}\leq0.035^{+0.003}_{-0.002},
\end{eqnarray}
where ``$=$" happens for $\delta^{\mathrm{lepton}}_{\mathrm{CP}}=0$
or $\delta^{\mathrm{lepton}}_{\rm CP}=90^\circ$. Though
$\delta^{\mathrm{lepton}}_{\rm KM}=90^\circ$ corresponds to
$\delta^{\mathrm{lepton}}_{\rm CK}=85.39^\circ$, there is a slight
difference in the $J^{\mathrm{lepton}}_{\rm CK}$ range if we let
$\delta^{\mathrm{lepton}}_{\rm CK}=90^\circ$. However, the
difference is of $\mathcal{O}(10^{-4})$ and we just neglect it.

It is interesting to notice that our procedure leads to a
quasi-maximal CP violation in the standard parametrization.
However, there have been some theoretical investigations indicating
that a large CP-violating phase $\delta^{\mathrm{lepton}}_{\rm CK}$
can be understood from some basic asymmetries. For example, the near
maximal CP violation with a large $\theta_{13}$ from our analysis is
in accordance with a general approach based on residual $Z_2$
symmetries~\cite{Ge:2011qn}. The maximal or large CP violation are
also predicted from some theoretical
reasonings~\cite{Wu:2012ri,He:2012yt,Hernandez:2012ra}, thus our
prediction of a quasi-maximal $\delta^{\mathrm{lepton}}_{\rm CK}$ or
a maximal $\delta^{\mathrm{lepton}}_{\rm KM}$ might acquire
theoretical support from basic considerations.

\section{\label{sec:moduli}THE SENSITIVITY OF THE MIXING MATRIX ELEMENTS TO THE CP-VIOLATING PHASE $\delta_{\rm CP}$}
Previous discussions are based on the hypothesis of the maximal CP
violation, while it is still helpful to check how the moduli of the
mixing matrix elements vary with the CP-violating phase $\delta_{\rm
CP}$. We can get the information on the sensitivity of $|V_{ij}|$
and $|U_{ij}|$ to $\delta_{\rm CP}$, and such information would be
useful when extracting the CP-violating information from the mixing
matrix elements. We make such an investigation in both the CK scheme
and the KM scheme.

\subsection{$|V_{ij}|$ as functions of $\delta_{\rm CK}$ and $\delta_{\rm KM}$}
Since the moduli of the CKM matrix elements are measured by various
processes, discussing the sensitivity of $|V_{ij}|$ to $\delta_{\rm
CP}$ can be useful when these measurements reach a higher precision
so that information on the CP violation can be extracted directly
from the moduli. We try to find out which elements are more
sensitive to the CP-violating phase.

\begin{figure*}
\includegraphics[origin=rb,width=0.6\textwidth]{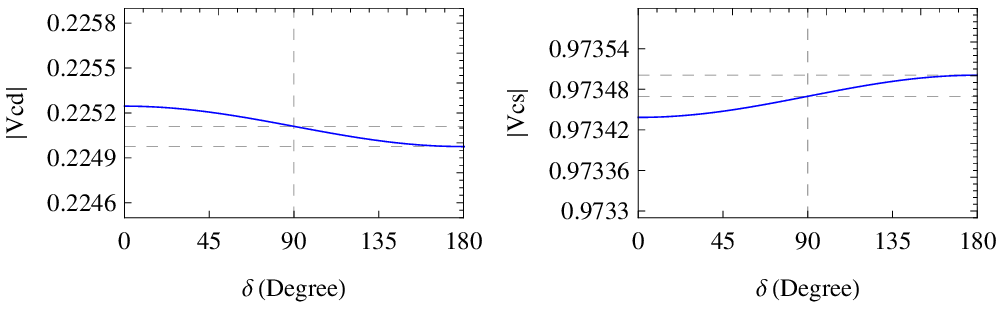}
\includegraphics[origin=lb,width=0.6\textwidth]{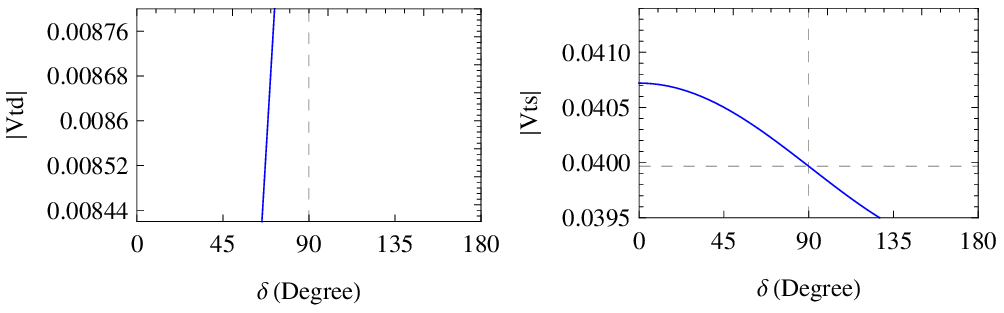}
\caption{\label{fig:ckv}$|V_{ij}|$ as a function of $\delta_{\rm
CK}$ in the CK scheme. The dashed lines denote $\delta_{\rm
CK}=90^\circ$ and $|V_{ij}|$ (for $\delta_{\rm
CK}=90^\circ, 180^\circ)$.}
\end{figure*}

The mixing angles in the angle-phase parametrizations are not
observables in the case of quark mixing, so we adopt the global fit
result of the Wolfenstein parameters to get the mixing angles in the
standard parametrization as we do in Section~\ref{sec:predict}. Thus
we get the CKM matrix with only one unrestrained parameter
$\delta_{\rm CK}$. Taking the range indicated by the global fit as a
reasonable one, we plot $|V_{ij}|$ as a function of $\delta_{\rm
CK}$ in Fig.~\ref{fig:ckv}.

\begin{figure*}
\includegraphics[origin=rb,width=0.6\textwidth]{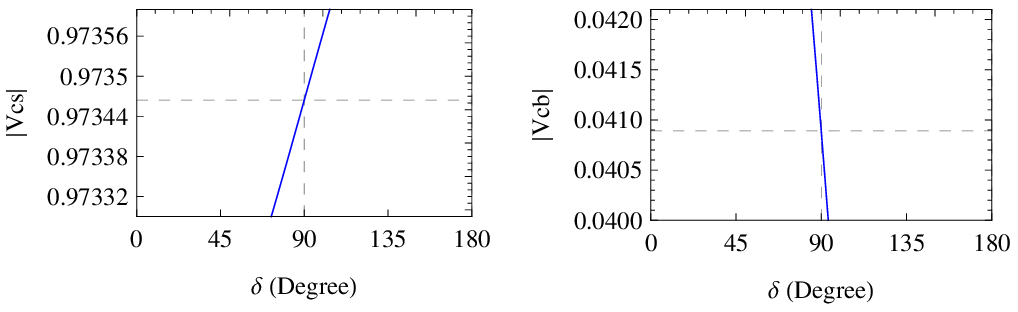}
\includegraphics[origin=lb,width=0.6\textwidth]{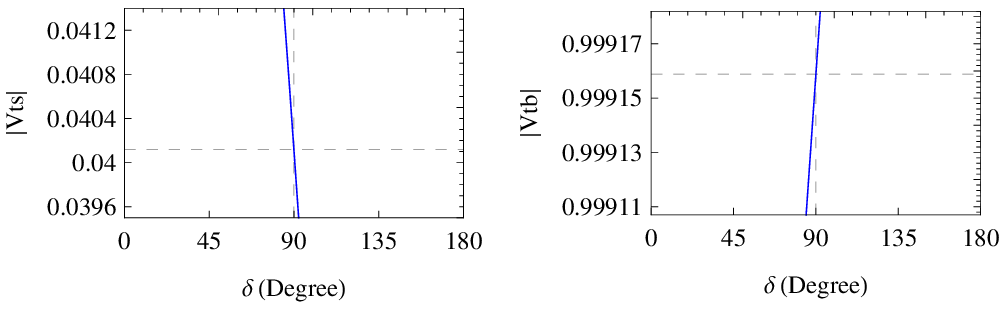}
\caption{\label{fig:kmv}$|V_{ij}|$ as a function of $\delta_{\rm
KM}$ in the KM scheme. The dashed lines denote $\delta_{\rm
CK}=90^\circ$ and $|V_{ij}|$ (for $\delta_{\rm
KM}=90^\circ, 180^\circ)$.}
\end{figure*}

Using the same procedure as in the previous section, namely,
calculating the mixing angles in the KM scheme with the matrix
elements that are independent of the phase, and leaving $\delta_{\rm
KM}$ unrestrained, we plot $|V_{ij}|$ as a function of $\delta_{\rm
KM}$ in Fig.~\ref{fig:kmv}.

From Fig.~\ref{fig:kmv}, we can see that in the range indicated by the global fit, the hypothesis of the maximal CP violation works quite well.

We list the range of $|V_{ij}|$ for a $\delta_{\rm CP}$ ranging $(0
- 180)^\circ$ in Table~\ref{tab:ckm2delta} for both the CK scheme
and the KM scheme. In each scheme, there are four $|V_{ij}|$
depending on $\delta_{\rm CP}$ and the four elements are not
necessarily the same with respect to their corresponding elements in
another scheme. With a certain scheme, which four elements are
dependent on $\delta_{\rm CP}$ is a question of phase convention. We
adopt the phase convention as in Eq.~[\ref{CK}] and Eq.~[\ref{KM}].
\begin{table*}
\caption{\label{tab:ckm2delta}The $\Delta|V_{ij}|$ range indicated by $\delta$ of $(0-180)^\circ$ }
\begin{ruledtabular}
 \begin{tabular}{ccc}
   \toprule
   $\Delta|V_{ij}|$ & The CK scheme & The KM scheme \\
   \hline
   $\Delta|V_{ud}|$ &-\footnote{``-" denotes this matrix element being independent of $\delta$.} & -\\
   $\Delta|V_{us}|$ &-      & -     \\
   $\Delta|V_{ub}|$ &-      & -     \\
   $\Delta|V_{cd}|$ &0.0003 & -     \\
   $\Delta|V_{cs}|$ &0.0001 &0.0011 \\
   $\Delta|V_{cb}|$ &-      &0.0292 \\
   $\Delta|V_{td}|$ &0.0066 & -     \\
   $\Delta|V_{ts}|$ &0.0015 &0.0300 \\
   $\Delta|V_{tb}|$ &-      &0.0011 \\
   \bottomrule
 \end{tabular}
\end{ruledtabular}
\end{table*}

From Table~\ref{tab:ckm2delta}, we see that, for $|V_{ij}|$ that are
dependent on different $\delta_{\rm CP}$, the dependence might
differ in orders of magnitude if we compare the dependence in terms
of $\Delta|V_{ij}|$ range, e.g., $\Delta|V_{cs}|$ is of
$\mathcal{O}(10^{-4})$ on dependence of $\delta_{\rm CK}$ and of
$\mathcal{O}(10^{-3})$ on dependence of $\delta_{\rm KM}$. More
explicitly, the dependence can be classified into,
\begin{enumerate}
\item $\mathcal{O}(10^{-2})$: $\Delta|V_{cb}|$, $\Delta|V_{ts}|$ versus $\delta_{\rm KM}$;
\item $\mathcal{O}(10^{-3})$: $\Delta|V_{td}|$, $\Delta|V_{ts}|$ versus $\delta_{\rm CK}$, and $\Delta|V_{cs}|$, $\Delta|V_{tb}|$ versus $\delta_{\rm KM}$;
\item $\mathcal{O}(10^{-4})$: $\Delta|V_{cd}|$, $\Delta|V_{cs}|$ versus $\delta_{\rm CK}$.
\end{enumerate}

From these results, we can observe that the KM scheme is more
sensitive when the CP-violating information is extracted from the
measured CKM matrix elements. The results also indicate that
$|V_{cb}|$ and/or $|V_{ts}|$ are good candidates for extracting
$\delta_{\rm KM}$.

\subsection{$|U_{ij}|$ as functions of $\delta_{\rm CK}$ and $\delta_{\rm KM}$}
\begin{figure*}
\includegraphics[origin=rb,width=0.6\textwidth]{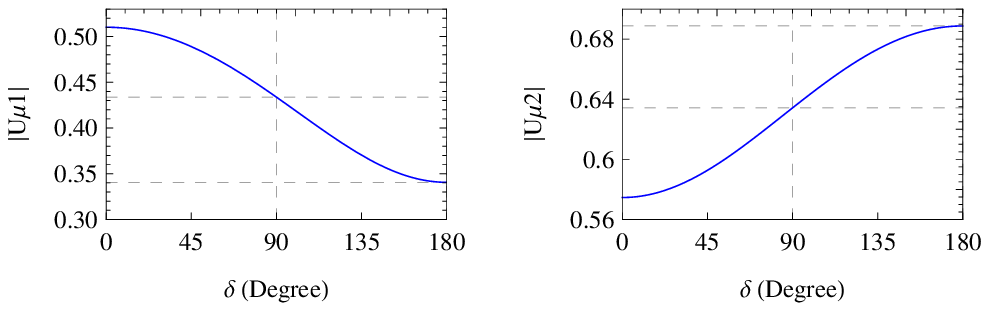}
\includegraphics[origin=lb,width=0.6\textwidth]{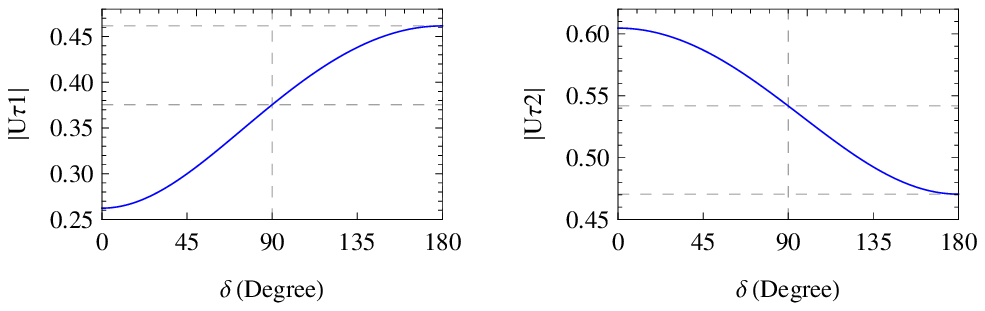}
\caption{\label{fig:cku}$|U_{ij}|$ as a function of $\delta_{\rm
CK}$ in the CK scheme. The dashed lines denote $\delta_{\rm
CK}=90^\circ$ and $|U_{ij}|$ (for $\delta_{\rm
CK}=90^\circ, 180^\circ)$.}
\end{figure*}

\begin{figure*}
\includegraphics[origin=rb,width=0.6\textwidth]{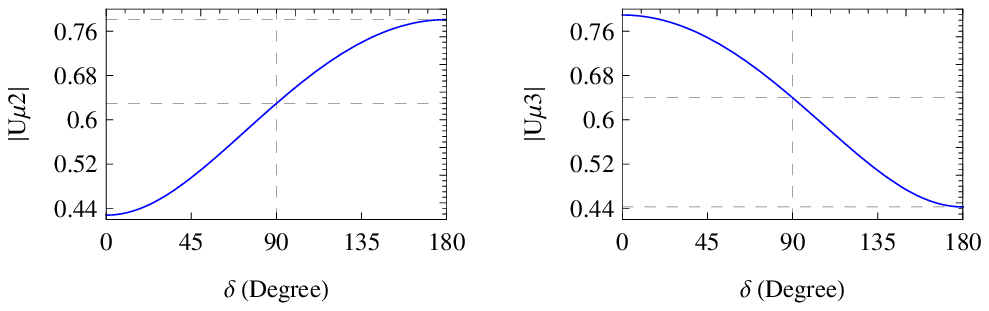}
\includegraphics[origin=lb,width=0.6\textwidth]{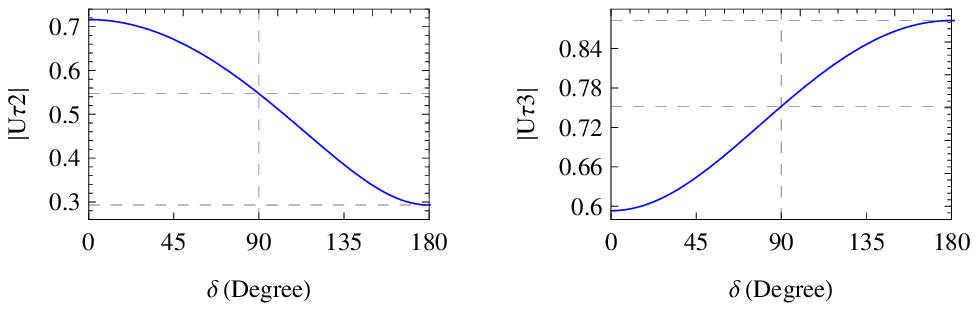}
\caption{\label{fig:kmu}$|U_{ij}|$ as a function of $\delta_{\rm
KM}$ in the KM scheme. The dashed lines denote $\delta_{\rm
KM}=90^\circ$ and $|U_{ij}|$ (for $\delta_{\rm
KM}=90^\circ, 180^\circ)$.}
\end{figure*}

The mixing angles in the neutrino mixing matrix are related to the
observed oscillation probabilities and are consequently observables.
We adopt the global fit values of $\theta_{12}$ and $\theta_{23}$,
and the averaged $\theta_{13}$ in Table~\ref{tab:theta13} as our
input. Since $|U_{ij}|$ are not well determined and there is no
experimental information on $\delta_{\rm CP}^{\mathrm{lepton}}$, we
plot $|U_{ij}|$ in a range where $\delta_{\rm CP}$ takes all its
possible values in Fig.~\ref{fig:cku} and Fig.~\ref{fig:kmu}.

We also calculate the $\Delta|U_{ij}|$ range and list the result in Table~\ref{tab:pmns2delta}.
\begin{table*}
\caption{\label{tab:pmns2delta} The $\Delta|U_{ij}|$ range indicated by $\delta$ of $(0-180)^\circ$ }
\begin{ruledtabular}
 \begin{tabular}{cccc}
   \toprule
   $\Delta|U_{ij}|$ & The CK scheme & The KM scheme \\
   \hline
   $\Delta|U_{e1}|$ &-\footnote{``-" denotes this matrix element being independent of $\delta$.} & -\\
   $\Delta|U_{e2}|$    &-      & -     \\
   $\Delta|U_{e3}|$    &-      & -     \\
   $\Delta|U_{\mu1}|$  &0.1695 & -     \\
   $\Delta|U_{\mu2}|$  &0.1142 &0.3532 \\
   $\Delta|U_{\mu3}|$  &-      &0.3465 \\
   $\Delta|U_{\tau1}|$ &0.1992 & -     \\
   $\Delta|U_{\tau2}|$ &0.1342 &0.4231 \\
   $\Delta|U_{\tau3}|$ &-      &0.2893 \\
   \bottomrule
 \end{tabular}
\end{ruledtabular}
\end{table*}
From Table~\ref{tab:pmns2delta}, we see that $\Delta |U_{ij}|$ are
all of the same order, which is different from the situation as in
quarks. Such a result indicates that all $|U_{ij}|$  within a
certain scheme are of similar sensitivity when they are used to
extract the CP-violating information. Besides, $\Delta |U_{ij}|$ is
larger for a variation of $\delta_{\rm KM}$. This indicates that
$|U_{ij}|$ are more sensitive to $\delta_{\rm KM}$.

It should be emphasized that the discussions here are only purposed
for illustration because of the ambiguity caused by the rephasing
transformation. Our discussions are based on the phase convention in
Eq.~[\ref{CK}] and Eq.~[\ref{KM}]. It is natural to seek for
phase-convention independent ways as we show in the following
section.

\section{\label{sec:phi} The MATRIX OF UNITARITY TRIANGLE ANGLES}
The physical observables are not affected by rephasing the
corresponding fields so it is better to find rephasing-invariant
descriptions. There have been several successful attempts. For
example, the Jarlskog invariant $J$, which was proposed by Jarlskog
in seeking for the commutator of the mass matrices~\cite{jarlskog},
is invariant under rephasing transformation. All CP violation
effects can be expressed as functions of $J$. It was also pointed
out by Wu in Ref.~\cite{Wu:1985ea} that special quartet forms like
$V_{j\beta}V_{k\gamma}(V_{j\gamma}V_{k\beta})^*$, ($i, j, k$ cyclic
and $\alpha, \beta, \gamma$ cyclic) are convention-invariant
quantities. For three generation case, there are nine such
quantities, which can form a matrix like,

\begin{eqnarray}
\Pi= \left(
  \begin{array}{ccc}
   V_{tb}V_{ts}^*V_{cs}V_{cb}^* &V_{td}V_{tb}^*V_{cb}V_{cd}^* &V_{ts}V_{td}^*V_{cd}V_{cs}^* \\
   V_{ub}V_{us}^*V_{ts}V_{tb}^* &V_{ud}V_{ub}^*V_{tb}V_{td}^* &V_{us}V_{ud}^*V_{td}V_{ts}^* \\
   V_{cb}V_{cs}^*V_{us}V_{ub}^* &V_{cd}V_{cb}^*V_{ub}V_{ud}^* &V_{cs}V_{cd}^*V_{ud}V_{us}^*
  \end{array} \right). ~~\label{Pi}
\end{eqnarray}

It was pointed out by Harrison, Dallison and Scott in
Ref.~\cite{Harrison:2009bz} that the matrix $-\Pi^*$ can be
decomposed into two matrices like,
\begin{eqnarray}
-\Pi^*= \left(
  \begin{array}{ccc}
   K_{ud}  & K_{us}   &  K_{ub}    \\
   K_{cd}  & K_{cs}   &  K_{cb}    \\
   K_{td}  & K_{ts}   &  K_{tb}
  \end{array} \right)+i\left(
  \begin{array}{ccc}
   J  & J   &  J    \\
   J  & J   &  J    \\
   J  & J   &  J
  \end{array} \right),
\end{eqnarray}
and we call the first one $K$ matrix. The orthogonality condition
aroused from the unitarity of the mixing matrix can be translated
into a geometrical language, i.e., the unitarity triangle. There are
six unitarity triangles for a 3 by 3 unitarity matrix. It is also
pointed out in Ref.~\cite{Harrison:2009bz} that the matrix of
unitarity triangle angles $\Phi$, can be constructed from the matrix
in Eq.~[\ref{Pi}],
\begin{eqnarray}
\Phi^{\mathrm{quark}}= \left(
  \begin{array}{ccc}
   \arg(-\Pi_{ud}^*)  &  \arg(-\Pi_{us}^*)   &  \arg(-\Pi_{ub}^*)    \\
   \arg(-\Pi_{cd}^*)  &  \arg(-\Pi_{cs}^*)   &  \arg(-\Pi_{cb}^*)    \\
   \arg(-\Pi_{td}^*)  &  \arg(-\Pi_{ts}^*)   &  \arg(-\Pi_{tb}^*)
  \end{array} \right),\label{Phi}
\end{eqnarray}
Each element in the $\Phi$ matrix corresponds to an inner angle of a
unitarity triangle, and 3 elements in each row or column correspond
to the three angles of a unitarity triangle. The $\Phi$ matrix is
rephasing-invariant, real and related to the geometrical image,
i.e., the unitarity triangles directly. The $K$ matrix is correlated
to the $\Phi$ matrix by $K=J\cot\Phi$. Similar $\Phi$ matrix can
also be constructed in the lepton sector.

Using the mixing angles and the phase indicated by Wolfenstein
parameters, we work out the complex CKM matrix in the CK scheme and
then get the following $\Phi^{\mathrm{quark}}$ matrix,
\begin{eqnarray}
\Phi^{\mathrm{quark}}_{\rm CK}= \left(
  \begin{array}{ccc}
   1.01^\circ    &  20.88^\circ   &  158.11^\circ \\
   67.83^\circ   &  90.31^\circ   &  21.86^\circ  \\
   111.16^\circ  &  68.81^\circ   &  0.03^\circ
  \end{array} \right).\label{ckPhi}
\end{eqnarray}
Similarly, we can construct the $\Phi^{\mathrm{quark}}$ matrix in
the KM scheme,
\begin{eqnarray}
\Phi^{\mathrm{quark}}_{\rm KM}= \left(
  \begin{array}{ccc}
   1.02^\circ    &  20.95^\circ   &  158.03^\circ \\
   68.10^\circ   &  89.97^\circ   &  21.93^\circ  \\
   110.88^\circ  &  69.08^\circ   &  0.03^\circ
  \end{array} \right). \label{kmPhi}
\end{eqnarray}
Note that the maximal CP violation hypothesis is used to get the
above result rather than deducing $\delta_{\rm KM}$ directly from
the Wolfenstein parameters.
Comparing these angles in Eq.~[\ref{ckPhi}] and Eq.~[\ref{kmPhi}],
we see that there are only slight difference in the obtained
$\Phi^{\mathrm{quark}}$ matrix elements between the two schemes.
Besides, in the $\Phi^{\mathrm{quark}}$ matrix, $\Phi_{us}$,
$\Phi_{cs}$, and $\Phi_{ts}$ correspond to $\beta$, $\alpha$, and
$\gamma$ in the $db$ unitarity triangle separately. We see that the
predicted $\beta$, $\alpha$, and $\gamma$ in the
$\Phi^{\mathrm{quark}}_{\rm KM}$ matrix are very close to the fit
values in the $\Phi^{\mathrm{quark}}_{\rm CK}$ matrix. Both values
for $\beta$, $\alpha$, and $\gamma$ are compatible with the measured
ones. This demonstrates that the maximal CP violation hypothesis
works well in an explicit way for quarks.

Taking the hypothesis of the maximal CP violation in the lepton sector, we can get the following $\Phi$ matrix,
\begin{eqnarray}
\Phi^{\mathrm{lepton}}_{\rm CK}= \left(
  \begin{array}{ccc}
   12.21^\circ    &  26.73^\circ   &  141.06^\circ \\
   78.36^\circ    &  78.95^\circ   &  22.69^\circ  \\
   89.44^\circ    &  74.32^\circ   &  16.25^\circ
  \end{array} \right),\label{ckPhi2}
\end{eqnarray}
where we use the predicted $\delta_{\rm CK}=85.38^\circ$ as an
input. Notice that this $\delta_{\rm CK}$ is deduced from a maximal
CP phase in the KM scheme, therefore we do not expect anything new
from the $\Phi^{\mathrm{lepton}}_{\rm KM}$ matrix except for very
small differences caused by precision.

From Eq.~[\ref{ckPhi}], Eq.~[\ref{kmPhi}], and Eq.~[\ref{ckPhi2}],
we see that $\sum_k\Phi_{ik}=180^\circ$ and
$\sum_i\Phi_{ij}=180^\circ$, i.e., the sum of every row and every
column of the $\Phi$ matrix equals to $180^\circ$ as they correspond
to three inner angles of a triangle.

\begin{figure*}
\includegraphics[width=0.8\textwidth]{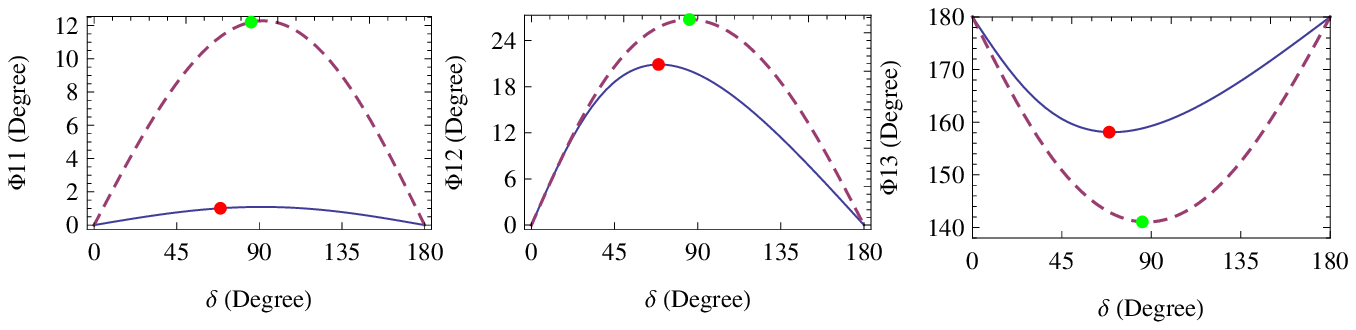}
\includegraphics[width=0.8\textwidth]{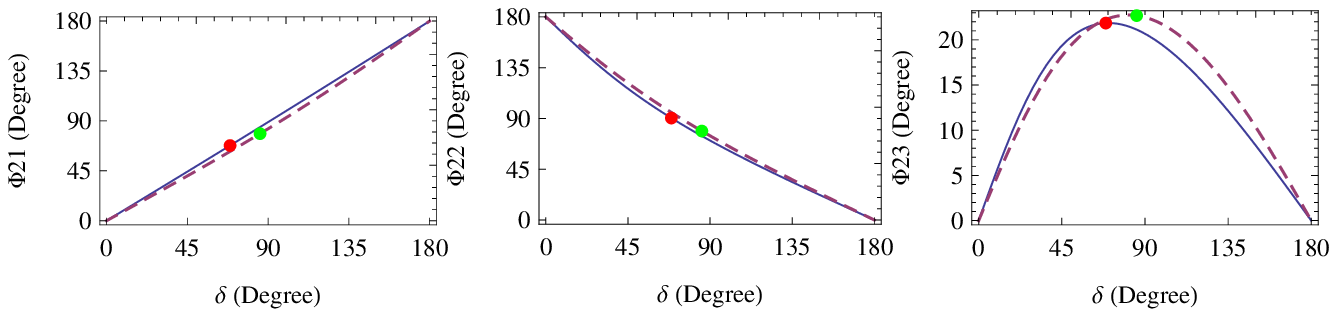}
\includegraphics[width=0.8\textwidth]{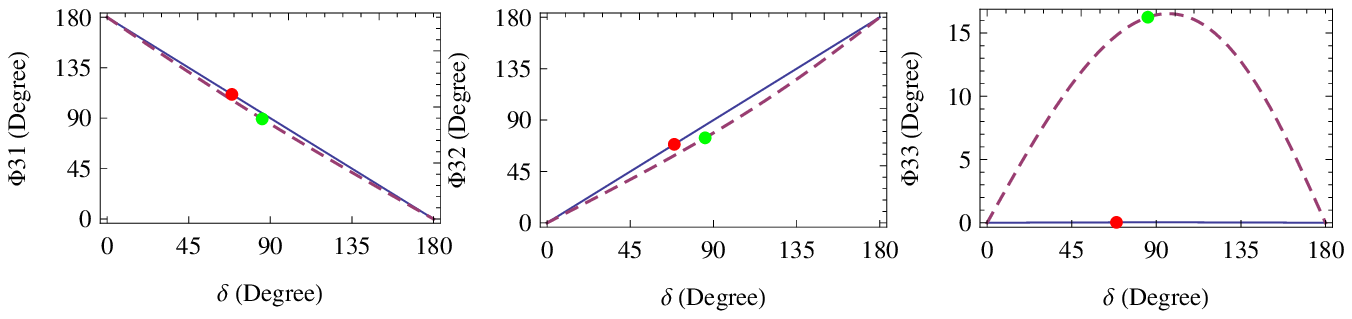}
\caption{\label{fig:ckphi}$\Phi_{ij}$ as a function of $\delta_{\rm
CP}$ in the CK scheme. The solid curve denotes
$\Phi_{ij}^{\mathrm{quark}}$; the dashed curve stands for
$\Phi_{ij}^{\mathrm{lepton}}$; the points correspond to $\delta_{\rm
CK}^{\mathrm{quark}}=68.8^\circ$ and $\delta_{\rm
CK}^{\mathrm{lepton}}=85.4^\circ$.}
\end{figure*}

\begin{figure*}
\includegraphics[width=0.8\textwidth]{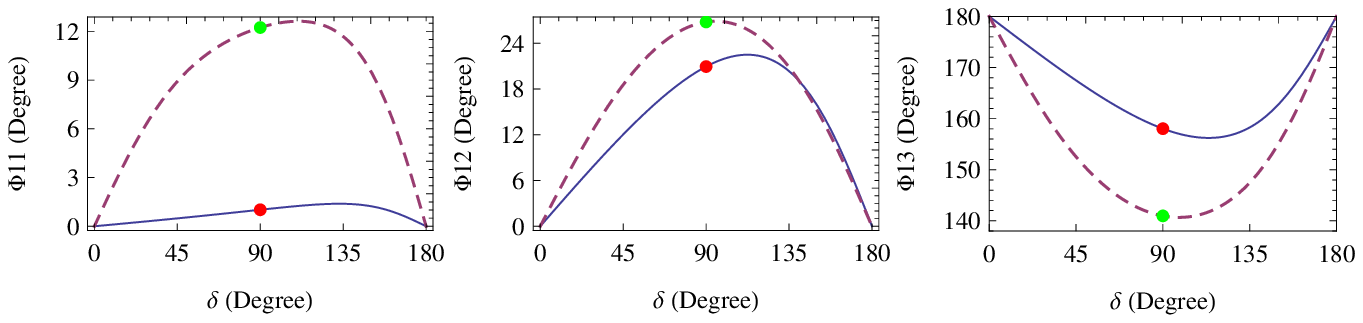}
\includegraphics[width=0.8\textwidth]{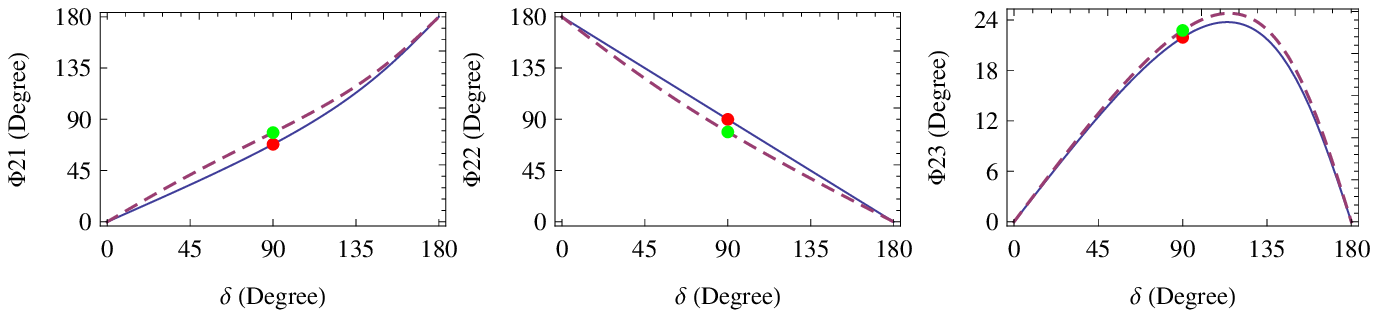}
\includegraphics[width=0.8\textwidth]{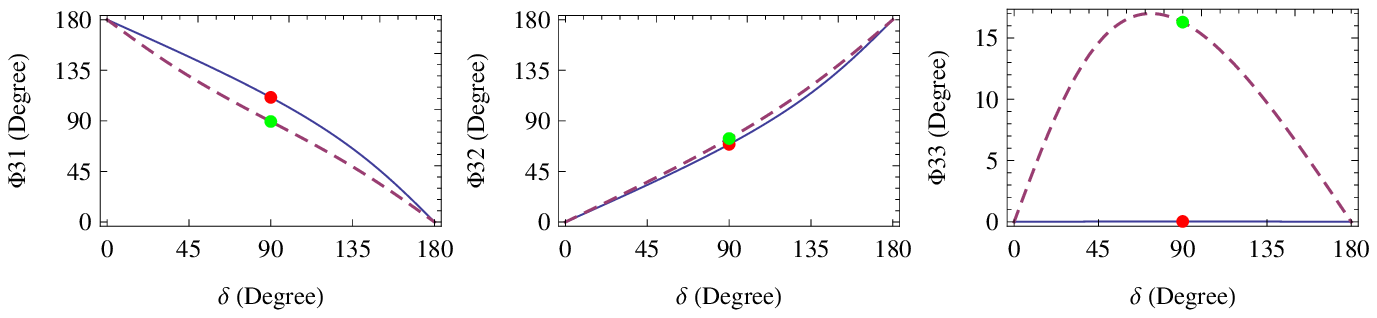}
\caption{\label{fig:kmphi}$\Phi_{ij}$ as a function of $\delta_{\rm
CP}$ in the KM scheme. The solid curve denotes
$\Phi_{ij}^{\mathrm{quark}}$; the dashed curve stands for
$\Phi_{ij}^{\mathrm{lepton}}$; the points correspond to $\delta_{\rm
KM}^{\mathrm{quark}}=90^\circ$ and $\delta_{\rm
KM}^{\mathrm{lepton}}=90^\circ$.}
\end{figure*}

Leaving $\delta_{\rm CP}$ unrestrained, we plot $\Phi_{ij}$ as
functions of $\delta_{\rm CK}$ and $\delta_{\rm KM}$ in
Fig.~\ref{fig:ckphi} and Fig.~\ref{fig:kmphi}.

From Fig.~\ref{fig:ckphi} and Fig.~\ref{fig:kmphi}, we find that,
\begin{enumerate}
\item the trends of $\Phi_{ij}$ with the variations of $\delta_{\rm CK}$ and $\delta_{\rm KM}$ are similar,
which indicates a weak dependence on parametrizations when other parameters
are fixed;
\item given same values for $\delta_{\rm CP}$, the lepton unitarity triangles are sizable, i.e., there is no
very small angles compared with those in the quark case;
\item the four elements in the left-bottom are sensitive to $\delta_{\rm CP}$ both in the quark and lepton mixing; these four elements can be in four unitarity triangles and the $db$ unitarity triangle in the quark mixing is one of the four;
\item $\Phi_{11}^{\mathrm{lepton}}$, $\Phi_{12}^{\mathrm{lepton}}$, $\Phi_{13}^{\mathrm{lepton}}$, $\Phi_{23}^{\mathrm{lepton}}$,
and $\Phi_{33}^{\mathrm{lepton}}$ are parabolic curves so two values
of $\delta$ are indistinguishable for only one element of
$\Phi_{ij}^{\mathrm{lepton}}$; in the CK scheme, these five elements
are almost symmetric versus $\delta_{\rm
CK}^{\mathrm{lepton}}=90^\circ$, which means that $\delta_{\rm
CK}^{\mathrm{lepton}}$ and $180^\circ-\delta_{\rm
CK}^{\mathrm{lepton}}$ are hard to be distinguished in these cases
for each $\Phi_{ij}^{\mathrm{lepton}}$.
\end{enumerate}

\section{\label{sec:sum}DISCUSSIONS AND CONCLUSIONS}

We have performed some investigations on the CP-violating phase
$\delta_{\rm CP}$ from several aspects.

In Section~\ref{sec:predict}, under the Ansatz of a maximal CP
violation $\delta_{\mathrm{KM}}^{\mathrm{lepton}}=90^\circ$ in the
lepton sector, we provide a prediction of the CP-violating phase
$\delta^{\mathrm{lepton}}_{\mathrm{CK}}=(85.39^{+4.76}_{-1.82})^\circ$
as well as a prediction of the PMNS matrix in
Eq.~[\ref{PMNSprediction}]. A replaying procedure is used firstly in
the quark sector as an exercise. We see that the hypothesis of the
maximal CP violation in the quark sector offers
$\delta^{\mathrm{quark}}_{\mathrm{CK}}=(68.62^{+0.89}_{-0.81})^\circ$,
which is close to the measured value
$\delta^{\mathrm{quark}}_{\mathrm{CK}}=(68.81^{+3.30}_{-2.17})^\circ$.
Besides, we may mention that the priori definition of maximal CP
violation, i.e., maximized $J$ under all four mixing parameters, has
been ruled out experimentally in the quark sector. In every
angle-phase parametrization $\delta_{\rm CP}$ shows up in
$\sin\delta_{\rm CP}$, so $\delta_{\rm CP}=90^\circ$ contributes
most to $J$ with respect of this parameter. We take this as our
definition of a maximal CP violation.

In Section~\ref{sec:moduli}, the variations of the moduli of the
mixing matrix elements with the CP-violating phase $\delta_{\rm CP}$
are demonstrated graphically. For $|V_{ij}|$ and $|U_{ij}|$ that are
dependent on $\delta_{\rm CP}$ under a certain scheme, we find that
$\Delta |V_{ij}|$ differ in orders of magnitude while $\Delta
|U_{ij}|$ are all of the same order. The KM scheme stands out as
more sensitive if the CP-violating information is extracted from the
measured elements of the CKM matrix or the PMNS matrix. We also find
that $|V_{cb}|$ and $|V_{ts}|$ are good candidates for extracting
$\delta_{\rm KM}^{\mathrm{quark}}$.  All the discussions in this
section are constrained to the phase convention in Eq.~[\ref{CK}]
and Eq.~[\ref{KM}].

The dependence of the matrix elements on $\delta_{\rm CP}$ is
convention-dependent, whereas physical observables are independent
of the phase convention. So we adopt the $\Phi$ matrix description
and make some discussions on it in Section~\ref{sec:phi}. In this
section, by using the convention-independent $\Phi$ matrix, we
continue our discussion by giving the $\Phi$ matrices for both
quarks and leptons under the Ansatz of a maximal CP violation. We
also demonstrate the variations of $\Phi_{ij}$ with
$\delta_{\mathrm{CP}}$ in both CK and KM schemes. We provide
predictions of all of the unitariy triangles which are directly
relevant to the CP violation effect in neutrino oscillation for
future experiments.

\begin{acknowledgments}
This work is partially supported by National Natural Science
Foundation of China (Grants No.~11021092, No.~10975003,
No.~11035003, and No.~11120101004) and by the Research Fund for the
Doctoral Program of Higher Education (China).
\end{acknowledgments}

\end{document}